\documentclass[onecolumn,draft, 11pt]{IEEEtran}

\usepackage{color}
\usepackage{graphicx,tabularx,array,amsmath,amsthm,thmtools}

\usepackage{mathtools}

\usepackage{amsfonts}
\usepackage{bm}
\usepackage{bbm}
\usepackage{makecell}
\usepackage{multirow}
 \usepackage{amssymb}
\usepackage{txfonts}
\usepackage[T1]{fontenc}
\usepackage{tikz}
\usepackage[scr=dutchcal]{mathalfa}
\usepackage{ textcomp }

\usepackage{cite}
\newcommand\blfootnote[1]{%
  \begingroup
  \renewcommand\thefootnote{}\footnote{#1}%
  \addtocounter{footnote}{-1}%
  \endgroup
}

\newtheorem{Theorem}{Theorem}
\newtheorem{Proposition}{Proposition}
\newtheorem{Lemma}{Lemma}
\newtheorem{Corollary}{Corollary}

\newtheorem{Remark}{Remark}
\newtheorem{Definition}{Definition}

\newcommand{\indep}{\rotatebox[origin=c]{90}{$\models$}}

\begin{document}

\title{A Concentration of Measure Approach to Database De-anonymization
}

\author{Farhad Shirani, Siddharth Garg  and Elza Erkip\\
Department of Electrical and Computer Engineering \\
New York University, NY. \\\date{} }

\maketitle
\begin{abstract}
In this paper, matching of correlated high-dimensional databases is investigated. A stochastic database model is considered where the correlation among the database entries is governed by an arbitrary joint distribution. Concentration of measure theorems such as typicality and laws of large numbers are used to develop a database matching scheme and derive necessary conditions for successful matching. Furthermore, it is shown that these conditions are tight through a converse result which characterizes a set of distributions on the database entries for which reliable matching is not possible. The necessary and sufficient conditions for reliable matching are evaluated in the cases when the database entries are independent and identically distributed as well as under Markovian database models.  \blfootnote{This work is supported by NYU WIRELESS Industrial Affiliates and
National Science Foundation grant CCF-1815821.} 
\end{abstract}

\section{Introduction}
The exponential growth in both storage and sharing of sensitive data in today's interconnected world has led to major privacy and security concerns. Databases containing \textit{micro-information} such as movie preferences, transaction data, and health records
are published and shared routinely in order to develop methods to improve recommendation systems, analyze financial markets, and facilitate research \cite{datta2012provable}. In order to preclude leakage of private information, the members' identities are often masked prior to publishing the database \cite{naini2016you,narayanan2008robust, takbiri2018matching}. However, it has been shown through several practical attack scenarios that such basic measures are insufficient in protecting the members' identities and privacy. Some of the well-known instances  of privacy breaches include the attack on the ostensibly anonymized Netflix prize database using publicly available data on the internet movie database (IMDB) \cite{narayanan2008robust}, the de-anonymization of a Massachusetts hospital discharge database using the cross-correlations with a public voter database \cite{sweeney1997weaving} and breaches caused
by the release of anonymized AOL search data \cite{NYT}. 

Despite the various practical attack algorithms proposed in the literature, a rigorous effort to investigate the conditions for successful matching has been missing until recently. In \cite{Negar_database_2018}, the authors take the first step in this direction and provide conditions for reliable matching by analyzing the performance of the maximum a-posteriori probability (MAP) algorithm for a general class of stochastically correlated databases. They provide conditions for the existence of successful matching algorithms as a function of the value of the cycle mutual information between the two databases. A related research direction uses differential privacy techniques to reduce  members' privacy risk by adding noise to database entries (in addition to anonymizing member identities) \cite{dwork2008differential}. The objective of differential privacy is publishing databases in which the effect of changes in a single entry element on the aggregated information in the database is negligible.
In \cite{takbiri2018matching}, the authors consider a relevant problem where the tradeoff between utility and privacy is investigated when database anonymization (removing member identities) and obfuscation (adding noise to database entries) is used to ensure privacy preservation. 

In this paper, we propose database de-anonymization (database matching) schemes and derive theoretical guarantees for successful de-anonymization. We consider a stochastic database model, where the database entries are correlated through a general joint distribution. 
Our formulation of the database matching problem defers from the one in \cite{Negar_database_2018} in several ways. First, in \cite{Negar_database_2018} it is required that a successful de-anonymization algorithm match all of the database entries correctly, whereas in this work we require that the fraction of entries which are matched correctly converge to one as the size of the database grows asymptotically. This relaxed criterion for successful matching plays a crucial role in the applicability of concentration of measure theorems used in this work.
Second, we consider a more general formulation where database entries may be generated based on discrete distributions or probability measures which are characterized by densities.  

We build upon our work on de-anonymization of graphical data \cite{shirani2018typicality,shirani2017seeded}, and fingerprinting de-anonymization attacks \cite{shirani2017information}, and use concentration of measure theorems to propose a \textit{typicality matching} scheme, where database entries are matched based on their joint typicality with respect to the underlying distribution. We leverage an extension of the Shannon-McMillan-Breiman theorem \cite{barron1985strong} to provide sufficient conditions for the success of the proposed scheme. In the next step, Fano's inequality \cite{cover2012elements} is used to provide tight necessary conditions for successful matching. We evaluate these results for two special classes of stochastic database models: i) \textit{I.I.D. database model} where the database entries are generated independently based on identical distributions, and ii) \textit{Markovian database model} where each database entry vector is generated based on a Markov random process.  The Markovian database model is of interest in various applications particularly those that model time-series data. For instance, health records databases follow the the Markovian stochastic model as the members' future health condition is related to its past through the present condition \cite{li2010section}.

The problem of database matching under the I.I.D. database model is closely related to that of matching of Erd\"{o}s-R\'{e}nyi graphs \cite{kazemi2015can}, where given a pair of stochastically correlated graphs, the objective is to find the canonical labeling of the vertices in the second graph based on the labeling of the first one. The pair of adjacency matrices of two Erd\"{o}s-R\'{e}nyi graphs resemble a pair of I.I.D. databases. However, there are \textit{fundamental} differences between the two matching problems. In graph matching, mislabeling a single vertex affects the adjacency matrix entries corresponding to all of the edges connected to that vertex \cite{shirani2018typicality}. Consequently, each mislabeled vertex results in a permutation of the adjacency matrix, whereas in database matching, mislabeling a member does not affect the database entries. Consequently, we are able to derive necessary and sufficient conditions for reliable database matching under general stochastic models which would not be possible in the graph matching problem.

In order to derive the aforementioned necessary and sufficient conditions for reliable database matching, we build upon the arguments used in classical information theory to characterize the capacity of point-to-point (PtP) channels with memory \cite{cover2012elements}. To this end, we construct an analogy between the `codebook' used in PtP channel coding problem and the labeled database in database matching. In this analogy, the labeled database entries are passed through a noisy test-channel to produce the unlabeled database entries. Successful matching of the database entries is analogous to reliable data transmission over the channel. While, the analogy between the channel coding problem and database matching is helpful in deriving necessary and sufficient conditions for successful matching, there are significant differences in the mathematical formulation of the problems. One main difference is that in channel coding, the codebook is designed to maximize the transmission rate, whereas in database matching, the database is given and cannot be modified to facilitate matching.

The rest of the paper is organized as follows:  Section \ref{Sec:Form} provides the formulation of the database matching problem. Section \ref{Sec:Prelim} describes the mathematical tools used in our analysis. Section \ref{Sec:Ach} explains the proposed matching scheme and provides sufficient conditions for successful matching. Section \ref{Sec:Con} provides necessary conditions for successful matching. Section \ref{Sec:Ex} evaluates these conditions under several stochastic models of interest. Section \ref{Sec:Conc} concludes the paper. 

\section{Notation}
\label{sec:notation}
 We represent random variables by capital letters such as $X, U$ and their realizations by small letters such as $x, u$. Sets and multisets are denoted by calligraphic letters such as $\mathcal{X}, \mathcal{U}$. The set of natural numbers, and the real numbers are shown by $\mathbb{N}$, and $\mathbb{R}$ respectively. The random variable $\mathbbm{1}_{\mathcal{E}}$ is the indicator function of the event $\mathcal{E}$. Random processes are shown by sans-serif letter $\mathsf{X},\mathsf{Y}$. For the random process $\mathsf{X}=(X_1,X_2,\cdots)$, the probability measure corresponding to the vector $X^n=(X_1,X_2,\cdots,X_n)$ is denoted by $P_{n,\mathsf{X}}$. The process $\mathsf{X}$ is said to be generated based on $P_{\mathsf{X}}= (P_{n,\mathsf{X}})_{n\in \mathbb{N}}$. 
 The set of numbers $\{1,2,\cdots, n\}, n\in \mathbb{N}$ is represented by $[n]$. 
 For a given $n\in \mathbb{N}$, the $n$-length vector $(x_1,x_2,\hdots, x_n)$ is written as $x^n$.

\section{Problem Formulation}
In this section, we provide the mathematical formulation of the database matching problem. A database consists of a set of \textit{entries}, each containing information corresponding to one of the \textit{members} of the database. An entry is a real-valued  vector generated based on a predetermined probability distribution.  
The following formally defines a database. 

\begin{Definition}[Deterministic Database]
An unlabeled database (UDB) is a multiset $\mathcal{D}_{m,n}=\{{u}^m_i\in \mathbb{R}^m| i\in [n]\}$, where $u^m_i=(u_{i,1},u_{i,2},\cdots, u_{i,m}), i\in [n]$ are called the entries of the database, $m\in \mathbb{N}$ is the length of the entries, and $n\in \mathbb{N}$ is the size of the database. A labeled database (LDB) $\overline{\mathcal{D}}_{m,n}$ is characterized by the pair $(\mathcal{D}_{m,n},\Theta)$, where the bijective mapping $\Theta: [n]\to [n]$ is called the labeling function.  The entry $u^m_i$ is said to correspond to the member indexed by $\Theta(i)$.
\end{Definition}

In this work, we study matching of randomly generated databases, when the size and length of the database grows asymptotically large. It is assumed that the information about the $i$th member is completely described by a random process $\mathsf{U}_{i'}= (U_{i',1},U_{i',2},\cdots)$, where $i=\Theta(i')$. The $i$th entry in a database with length $m$ consists of the first $m$ random variables in the random process. The random variables in each database entry - which correspond to the same member - may be correlated with each other. However, entries corresponding to different members are assumed to be generated independently of each other.
As a result, a randomly generated database is completely described by its size $m$, entry length $n$, and the underlying distribution for each of its entries $P_{\mathsf{U}_i}, i\in [n]$. This is formalized below.

\begin{Definition}[Random Database]
An $(m,n,\{{P}_{\mathsf{U}_i}\}_{i\in [n]})$-UDB is a stochastically generated multiset $\mathcal{D}_{m,n}=\{{U}^m_i\in \mathbb{R}^m| i\in [n]\}$, where $U^m_i=(U_{i,1},U_{i,2},\cdots, U_{i,m}), i\in [n]$ consist of the first $m$ variables in the stationary random processes $\mathsf{U}_i, i\in [n]$ which is generated according to $P_{\mathsf{U}_i}$. An $(m,n,\{{P}_{\mathsf{U}_i}\}_{i\in [n]})$-LDB $\overline{\mathcal{D}}_{m,n}=(\mathcal{D}_{m,n},\Theta)$ consists of an  $(m,n,\{{P}_{\mathsf{U}_i}\}_{i\in [n]})$-UDB $\mathcal{D}_{m,n}$ and a labeling function $\Theta$. 
\end{Definition}

We consider pairs of correlated databases. It is assumed that the two databases have equal size and their entries correspond to the same set of members. Pairs of entries corresponding to the same member are called matching entries. These entries are correlated with each other and are generated based on a joint distribution and independent of all other entries.

\begin{Definition}[Correlated Databases]
A pair of correlated databases (CLDB) is a pair $(\overline{\mathcal{D}}^{(1)}_{m,n},\overline{\mathcal{D}}^{(2)}_{m,n})$, where $\overline{\mathcal{D}}^{(j)}_{m,n}=(\mathcal{D}^{(j)}_{m,n},\Theta^{(j)}), j\in \{1,2\}$, and  ${\mathcal{D}}^{(j)}_{m,n}=\{{U}^{(j),m}_i\in \mathbb{R}^m| i\in [n]\}$ is an $(m,n,\{{P}_{\mathsf{U}^{(j)}_i}\}_{i\in [n]})$-UDB. Two entries $U^{(1),m}_{i_1}$ and $U^{(2),m}_{i_2}$ are called matching if $\Theta^{(1)}(i_1)=\Theta^{(2)}(i_2)$. The matching entries $U^{(1),m}_{i_1}$ and $U^{(2),m}_{i_2}$ are a pair of correlated processes generated according to the joint distribution $P_{\mathsf{U}^{(1)}_{i_1},\mathsf{U}^{(2)}_{i_2}}$. Entries which are not matching are generated independently of each other.
\end{Definition}

\begin{Remark}
For brevity, we assume that the entries corresponding to distinct members in the pair of correlated databases are distributed identically and independently of each other. More precisely, we assume that $P_{\mathsf{U}^{(1)},\mathsf{U}^{(2)}}= P_{\mathsf{U}^{(1)}_i,\mathsf{U}^{(2)}_i}, \forall i\in [n]$. In this case, a CLDB is completely characterized by the tuple $(m,n,\Theta^{(1)}, \Theta^{(2)}, P_{\mathsf{U}^{(1)},\mathsf{U}^{(2)}})$.
It is straightforward to extend the results presented in this work to the case when the database entries are generated based on distributions which are not identical with each other.  
\end{Remark}

The objective in the database matching problem is to leverage the correlation among the entries of two stochastically correlated databases to match the labels of their members. We consider families of databases whose entry length $m$ and member-set size $n_m$ grows asymptotically large, where
the value $R=\lim_{m\to \infty}\frac{1}{m}\log_2{n_m}$ is called the \textit{rate of growth} of the database. For a given $m,n_m\in \mathbb{N}$, a matching scheme takes the pair $(\overline{\mathcal{D}}^{(1)}_{m,n_m},{\mathcal{D}}^{(2)}_{m,n_m})$ as its input, where the labeling function for the first database is given, whereas the second labeling function is missing. The scheme outputs a reconstruction of the labeling function for ${\mathcal{D}}^{(2)}_{m,n_m}$. The scheme is said to be successful if the fraction of members which are matched correctly approaches one as the length $m$ and size $n_m$ of the database is increased asymptotically.

\begin{Definition}[Family of Databases]
A family of CLDBs $(\overline{\mathcal{D}}^{(1)}_{m,n_m},\overline{\mathcal{D}}^{(2)}_{m,n_m}), m\in \mathbb{N}$ is a sequence of databases generated according to $P_{\mathsf{U}^{1},\mathsf{U}^{2} }$, where $(n_m)_{m\in \mathbb{N}}$ is an increasing sequence of natural numbers. Each labeled database $\overline{\mathcal{D}}^{(j)}_{m,n_m}$ consists of the pair $({\mathcal{D}}^{(j)}_{m,n_m}, \Theta^{(j)}_m), j\in \{1,2\}$, where ${\mathcal{D}}^{(j)}_{m,n_m}=\{{U}^{(j),m}_i\in \mathbb{R}^m| i\in [n]\}$, and $ U^{(j),m}_I= (U^{(j)}_{I,1}, U^{(j)}_{I,2}, \cdots, U^{(j)}_{I,m})$. It is assumed that $\Theta^{(j)}_{m}(i)= \Theta^{(j)}_{m'}(i), \forall i\leq min(m,m'), m,m'\in \mathbb{N}$. The family of CLDBs is characterized by the tuple $((n_m)_{m\in \mathbb{N}},(\Theta^{(1)}_m)_{m\in \mathbb{N}}, (\Theta^{(2)}_m)_{m\in \mathbb{N}}, P_{\mathsf{U}^{(1)},\mathsf{U}^{(2)}})$. The rate of growth (rate) of the database is defined as $R= \lim_{m\to \infty} \frac{1}{m}\log_2{n_m}$.
\end{Definition}

In this work, we consider families of CLDBs whose rate of growth is finite.

\begin{Definition}[Database Matching Algorithm]
Consider a family of CLDBs  $(\overline{\mathcal{D}}^{(1)}_{m,n_m},\overline{\mathcal{D}}^{(2)}_{m,n_m}), m\in \mathbb{N}$ characterized by  $((n_m)_{m\in \mathbb{N}},(\Theta^{(1)}_m)_{m\in \mathbb{N}}, (\Theta^{(2)}_m)_{m\in \mathbb{N}}, P_{\mathsf{U}^{(1)},\mathsf{U}^{(2)}})$. A matching scheme is a sequence of mappings $f_m: (\overline{\mathcal{D}}^{(1)}_{m,n_m},{\mathcal{D}}^{(2)}_{m,n_m})\to \widehat{\Theta}_m^{(2)}$. The scheme is called a successful matching scheme if 
\begin{align}
\label{eq:match}
P(\Theta^{(2)}_m(U^{(2),m}_{I})=\widehat{\Theta}^{(2)}_m(U^{(2),m}_{I})) \to 1 \text{ as } m\to \infty,
\end{align}
where $I$ is uniformly distributed over $[n_m]$.
\end{Definition}

It can be noted that the criteria for successful matching in Equation \eqref{eq:match} requires the fraction of members which have been matched correctly to go to one as the size and length of the database grow asymptotically. This is in contrast with \cite{Negar_database_2018} where all of the database entries are required to be matched correctly simultaneously. The relaxation allows us to use the \textit{typicality matching scheme} which is described in the next sections. 
Our objective is to find the matchability region $\mathcal{R}$, that is, the set of $(R, P_{\mathsf{U}^{(1)},\mathsf{U}^{(2)}})$ pairs for which a successful matching algorithm exists. 

\begin{Definition}[Matchability Region]
The pair $(R, P_{\mathsf{U}^{(1)},\mathsf{U}^{(2)}})$ is said to be matchable, if for any family of CLDBs with rate of growth $R$ generated according to $P_{\mathsf{U}^{(1)},\mathsf{U}^{(2)}}$, there exists a successful matching scheme. The set of all matchable $(R, P_{\mathsf{U}^{(1)},\mathsf{U}^{(2)}})$ pairs is called the matchability region and is denoted by $\mathcal{R}$. 
\end{Definition}

\label{Sec:Form}

\section{Preliminaries}
\label{Sec:Prelim}
In our derivations, we make use of the asymptotic equipartition property (AEP)
of random processes. The AEP was first proved for finite-valued, stationary and ergodic processes using the Shannon-McMillan-Breiman theorem \cite{breiman1957individual}. An extension of this theorem was later proved for stationary and ergodic random processes which take values over a standard Borel space \cite{barron1985strong}. In this section, we provide a brief summary of the latter result and its relevant implications. 

Consider a stochastic process $\mathsf{X}=(X_1,X_2,\cdots)$, where $X_i$ take values over a standard Borel space. Let the joint distribution on the first $n$ elements in the process $(X_1,X_2,\cdots,X_n)$ be denoted by $P_n$. Assume that $P_n$ is absolutely continuous with respect to the Lebesgue measure so that the density function $f_n(\cdot)=f_{X_1,X_2,\cdots,X_n}(\cdot)$ exists. Conditional densities $g_{n+1}(\cdot|\cdot)=f_{X_{n+1}| X_1,X_2,\cdots,X_n}(\cdot|\cdot), n\in \mathbb{N}$ are defined in the standard way. 
\begin{Definition}[Entropy]
For the random process $\mathsf{X}=(X_1,X_2,\cdots)$ characterized by the family of densities $f_{X_1,X_2,\cdots,X_n}, n\in \mathbb{N}$, the relative entropy rate is defined as:
\begin{align*}
    H_{RER}(\mathsf{X})= \lim_{n\to \infty} \mathbb{E}(-\log{[g_{n+1}(X_{n+1}|X_1,X_2,\cdots,X_n)]}).
\end{align*}
\end{Definition}

\begin{Lemma}[Barron \cite{barron1985strong}]
\label{Lem:SMB}
For the random process $\mathsf{X}=(X_1,X_2,\cdots)$ characterized by the family of densities $f_{X_1,X_2,\cdots,X_n}, n\in \mathbb{N}$, the following holds:
\begin{align*}
    -\frac{1}{n}\log [f_n(X_1,X_2,\cdots,X_n)]
    \to H_{RER}(\mathsf{X}),
\end{align*}
where the convergence is in the almost sure sense. 
\end{Lemma}
The following is a direct consequence of Lemma \ref{Lem:SMB}.

\begin{Proposition}[Typicality]
\label{Prop:Typ}
Define the typical set associated with the random process $\mathsf{X}$ as:
\[\mathcal{A}^n_{\epsilon}(\mathsf{X})= 
\{x^n\big| |-\frac{1}{n}\log{[f_n(x^n)]}- H_{RER}(\mathsf{X})|\leq \epsilon\},
\]
where $\epsilon>0$, and $n\in \mathbb{N}$. Then, 
\begin{enumerate}
    \item $P(\mathcal{A}^n_\epsilon)(\mathsf{X})\to 1$ as $n\to \infty$.
    \item $x^n\in \mathcal{A}_\epsilon^n(\mathsf{X})\Rightarrow 
    2^{-n(H_{RER}(\mathsf{X})+\epsilon)}\leq f_n(x^n)\leq 2^{-n(H_{RER}(\mathsf{X})-\epsilon)}$
    \item $2^{n(H_{RER}(\mathsf{X})-\epsilon)} \leq \mathcal{M}(\mathcal{A}_\epsilon^n(\mathsf{X}))\leq 2^{n(H_{RER}(\mathsf{X})+\epsilon)} $ for large enough $n$, where $\mathcal{M}(\cdot)$ is the Lebesgue measure. 
\end{enumerate} 

\end{Proposition}

The AEP holds for pairs of correlated stochastic processes as well. Consider the pair of processes $(\mathsf{X},\mathsf{Y})$ characterized by the joint density function $f_{n,\mathsf{X},\mathsf{Y}}(\cdot)= f_{X^n,Y^n}(\cdot)$, where the density is written with respect to the Lebesgue measure. The following is a multivariate extension of Proposition \ref{Prop:Typ}.  

\begin{Proposition}[Joint Typicality]
\label{Prop:Jtyp}
Define the typical set associated with the random processes $(\mathsf{X},\mathsf{Y})$ as:
\[\mathcal{A}^n_{\epsilon}(\mathsf{X},\mathsf{Y})= 
\{(x^n,y^n)\big| |-\frac{1}{n}\log{[f_{n,\mathsf{X},\mathsf{Y}}(x^n,y^n)]}- H_{RER}(\mathsf{X},\mathsf{Y})|\leq \epsilon\},
\]
where $\epsilon>0$, and $n\in \mathbb{N}$. Then, 
\begin{enumerate}
    \item $P(\mathcal{A}^n_\epsilon(\mathsf{X},\mathsf{Y}))\to 1$ as $n\to \infty$.
    \item $(x^n,y^n)\in \mathcal{A}_\epsilon^n(\mathsf{X},\mathsf{Y})\Rightarrow $\\
    $2^{-n(H_{RER}(\mathsf{X},\mathsf{Y})+\epsilon)}\leq f_n(x^n,y^n)\leq 2^{-n(H_{RER}(\mathsf{X},\mathsf{Y})-\epsilon)}$
    \item $2^{n(H_{RER}(\mathsf{X},\mathsf{Y})-\epsilon)} \leq \mathcal{M}(\mathcal{A}_\epsilon^n(\mathsf{X},\mathsf{Y}))\leq 2^{n(H_{RER}(\mathsf{X},\mathsf{Y})+\epsilon)}$ for large enough $n$. 
\end{enumerate} 
\end{Proposition}

We use the following result which follows from Propositions \ref{Prop:Typ} and \ref{Prop:Jtyp} using standard information theoretic arguments. 

\begin{Proposition}
\label{Prop:Indep}
Consider the correlated pair of stochastic processes $(\mathsf{X},\mathsf{Y})$ characterized by the sequence of joint densities $f_{n,\mathsf{X},\mathsf{Y}}, n\in \mathbb{N}$. Let $f_{n,\mathsf{X}}$ and $f_{n,\mathsf{Y}}$ be the marginal densities corresponding to $\mathsf{X}$ and $\mathsf{Y}$, respectively. Assume that the processes $\mathsf{X}'$ and $\mathsf{Y}'$ are generated according to the marginals $f_{n,\mathsf{X}}$ and $f_{n,\mathsf{Y}}$ independently of each other. Then,
\begin{align*}
    P(({X'}^n,{Y'}^n)\in \mathcal{A}^n_{\epsilon}(\mathsf{X},\mathsf{Y}))\leq 
    2^{-n(I(\mathsf{X};\mathsf{Y})-3\epsilon)},
\end{align*}
for large enough $n$, where the mutual information is defined as $I(\mathsf{X};\mathsf{Y})=H_{RER}(\mathsf{X})+H_{RER}(\mathsf{X})- H_{RER}(\mathsf{X},\mathsf{Y})$, $n\in \mathbb{N}$ and $\epsilon>0$.

\end{Proposition}

\section{Database Matching Scheme}
\label{Sec:Ach}
In this section, we propose a database matching scheme based on the concept of joint typicality of stochastic processes described in the previous section. Recall that in the problem under consideration, we are given a pair of correlated databases along with the labeling function for the first database. The objective is to find a faithful reconstruction of the labeling function of the second database. To this end, for each entry in the second database, the matching scheme finds a unique entry in the first database which for which the two entries are jointly typical. If such a unique entry exists, then the two entries are matched. Otherwise, the entry is added to an ambiguity set. Once this process is performed for all database entries, the unmatched entries in the ambiguity set are matched using a random and uniform index assignment function. This is described in more detail in the following.

Fix $m,n_m\in \mathbb{N}$, $\epsilon>0$, and $P_{\mathsf{U}^{(1)},\mathsf{U}^{(2)}}$. Let $(\overline{\mathcal{D}}^{(1)}_{m,n},\overline{\mathcal{D}}^{(2)}_{m,n})$ be a CLDB and assume that we are given $(\overline{\mathcal{D}}^{(1)}_{m,n},{\mathcal{D}}^{(2)}_{m,n})$. The matching scheme finds $\widehat{\Theta}^{(2)}(i'), i'\in [n_m]$ which is the reconstruction of the value of the labeling function for the $i'$th entry in the second database as follows. If there exists a unique entry $U^{(1),m}_{i}\in {\mathcal{D}}^{(i')}_{m,n}, i\in [n_m]$ such that 
\begin{align*}
    (U^{(1),m}_{i},U^{(2),m}_{i'})\in \mathcal{A}^m_{\epsilon}(\mathsf{U}^{(1)},\mathsf{U}^{(2)}),
\end{align*}
then $\widehat{\Theta}^{(2)}(i')= \Theta^{(1)}(i)$, otherwise the index $i'$ is added to the ambiguity set $\mathcal{L}'$. Define the following set:

\begin{align*}
    \mathcal{L}= \{i| i\in [n_m]- Im(\widehat{\Theta}^{(2)})\},
\end{align*}
where $Im(f)$ is the image of $f$. The values $\widehat{\Theta}^{(2)}(i'), i'\in \mathcal{L}'$ are then chosen randomly, uniformly and without replacement from the set $\mathcal{L}$. We call this scheme the \textit{typicality matching} scheme. 
\begin{Theorem}
\label{th:1}
For a family of CLDBs  $(\overline{\mathcal{D}}^{(1)}_{m,n_m},\overline{\mathcal{D}}^{(2)}_{m,n_m}), m\in \mathbb{N}$ characterized by  $((n_m)_{m\in \mathbb{N}},(\Theta^{(1)}_m)_{m\in \mathbb{N}}, (\Theta^{(2)}_m)_{m\in \mathbb{N}}, P_{\mathsf{U}^{(1)},\mathsf{U}^{(2)}})$, let $R$ be the rate of growth of the database. The typicality matching scheme is successful if the following conditions hold:
\begin{align*}
    R< I(\mathsf{U}^{(1)};\mathsf{U}^{(2)}).
\end{align*}
\end{Theorem}
\textit{Outline of proof.} Consider the matching process for the $i'$th entry. Define the following two error events:
\begin{align*}
 & \mathcal{E}_{i',1}: \nexists  i\ni (U^{(1),m}_{i},U^{(2),m}_{i'})\in \mathcal{A}^m_{\epsilon}(\mathsf{U}^{(1)},\mathsf{U}^{(2)})
    \\& \mathcal{E}_{i',2}: \exists i''\ni (U^{(1),m}_{i''},U^{(2),m}_{i'})\in \mathcal{A}^m_{\epsilon}(\mathsf{U}^{(1)},\mathsf{U}^{(2)})
    \\& \qquad \qquad\& \quad \Theta^{(1)}(i'') \neq \Theta^{(2)}(i')
\end{align*}
The event $\mathcal{E}_{i',1}$ is the event that a pair of jointly typical entries does not exit in the two databases, and the event $\mathcal{E}_{i',2}$ is the event that either the jointly typical pair is not unique or that the pair is not a matching pair. Let $\mathcal{E}_{i'}$ be the probability of a mismatch for the $i'$th entry. It is straightforward to show that $P(\mathcal{E}_{i'}) \leq P(\mathcal{E}_{i',1}\cup \mathcal{E}_{i',2}) = P(\mathcal{E}_{i',1})+ P(\mathcal{E}_{i',2}|\mathcal{E}^c_{i',1})$. Using Propositions \ref{Prop:Jtyp} and \ref{Prop:Indep} along with standard information theoretic arguments, we have:
\begin{align*}
    &P(\mathcal{E}_{i',1})\to 0 \text{ as } m\to \infty,
    \\&P(\mathcal{E}_{i',2}|\mathcal{E}^c_{i',1}) \leq 2^{mR}\cdot 2^{m(I(\mathsf{U}^{(1)};\mathsf{U}^{(2)})-3\epsilon)} \to 0 \text{ as } m\to \infty,
\end{align*}
where convergence in the second equation follows from the assumption $R< I(\mathsf{U}^{(1)};\mathsf{U}^{(2)})$ and taking $\epsilon$ to be small enough. 

\section{Converse}
\label{Sec:Con}
In this section, we derive necessary conditions for the existence of successful matching schemes. The following theorem states the main result of this section. 
\begin{Theorem}
\label{th:2}
Let $(\Theta_m^{(1)},\Theta_m^{(2)}), m\in \mathbb{N}$ be a family of pairs of labeling functions chosen randomly and uniformly over the set of all labeling functions. For a family of CLDBs  $(\overline{\mathcal{D}}^{(1)}_{m,n_m},\overline{\mathcal{D}}^{(2)}_{m,n_m}), m\in \mathbb{N}$ characterized by  $((n_m)_{m\in \mathbb{N}},(\Theta^{(1)}_m)_{m\in \mathbb{N}}, (\Theta^{(2)}_m)_{m\in \mathbb{N}}, P_{\mathsf{U}^{(1)},\mathsf{U}^{(2)}})$, let $R$ be the rate of growth of the database. Then, a necessary condition for the existence of a successful matching scheme is:
\begin{align*}
    R\leq I(\mathsf{U}^{(1)};\mathsf{U}^{(2)}).
\end{align*}
\end{Theorem}
\textit{Outline of the proof.} Let $P_e$ be the probability of the scheme being unsuccessful. Then, from Fano's inequality we have:
\begin{align*}
    \frac{1}{n_m}H(\Theta_m^{(2)}|\overline{\mathcal{D}}^{(1)}_{m,n_m},\mathcal{D}^{(2)}_{m,n_m})
    &\leq \frac{1}{n_m}+ \frac{1}{n_m}P_e\log{(n_m!)}
   \\    &\leq \frac{1}{n_m}+ P_e\log{n_m},
\end{align*}
where in the last inequality we have used the fact that $k!\leq k^k$. As a result,
\begin{align*}
    \frac{1}{n_m}H(\Theta^{(2)})
    &= \frac{1}{n_m}H(\Theta_m^{(2)}|\overline{\mathcal{D}}^{(1)}_{m,n_m},\mathcal{D}^{(2)}_{m,n_m})+ \frac{1}{n_m}I(\Theta_m^{(2)};\overline{\mathcal{D}}^{(1)}_{m,n_m},\mathcal{D}^{(2)}_{m,n_m})
    \\&\leq  \frac{1}{n_m}+ P_e\log{n_m}+\frac{1}{n_m}I(\Theta_m^{(2)};\overline{\mathcal{D}}^{(1)}_{m,n_m},\mathcal{D}^{(2)}_{m,n_m}).
\end{align*}
On the other hand:
\begin{align*}
    &I(\Theta_m^{(2)};\overline{\mathcal{D}}^{(1)}_{m,n_m},\mathcal{D}^{(2)}_{m,n_m})
    = I(\Theta_m^{(2)};{\mathcal{D}}^{(2)}_{m,n_m})+
    I(\Theta_m^{(2)};\overline{\mathcal{D}}^{(1)}_{m,n_m}|\mathcal{D}^{(2)}_{m,n_m})
    \\&\stackrel{(a)}{=}I(\Theta_m^{(2)};\overline{\mathcal{D}}^{(1)}_{m,n_m}|\mathcal{D}^{(2)}_{m,n_m})
    =I(\Theta_m^{(2)},\mathcal{D}^{(2)}_{m,n_m};\overline{\mathcal{D}}^{(1)}_{m,n_m})
    \\&= I(\overline{\mathcal{D}}^{(2)}_{m,n_m};\overline{\mathcal{D}}^{(1)}_{m,n_m})
    \stackrel{(b)}{=} \sum_{i\in [n_m]} I(U^{(1),m}_{{\Theta^{(1)}}^{(-1)}(i)};U^{(2),m}_{{\Theta^{(2)}}^{(-1)}(i)})
    \\&\leq n_m mI(\mathsf{U}_1;\mathsf{U}_2),
\end{align*}
where (a) follows from $\Theta_m^{(2)} \indep {\mathcal{D}}^{(2)}_{m,n_m}$ and (b) follows from the fact that entries which are not matching are generated independently of each other. So far, we have shown:
\begin{align*}
   \frac{1}{n_m}H(\Theta^{(2)})\leq \frac{1}{n_m}+ P_e\log{n_m}+ mI(\mathsf{U}_1;\mathsf{U}_2).
\end{align*}
On the other hand, we have $H(\Theta^{(2)}) \to n_m\log{n_m}$ as $n_m\to \infty$. Hence, as $n_m\to \infty$ we must have, 
\begin{align*}
   &\log{n_m}\leq  P_e\log{n_m}+ mI(\mathsf{U}_1;\mathsf{U}_2)\\
   &\Rightarrow (1-P_e)(\frac{1}{m}\log{n_m})\leq I(\mathsf{U}_1;\mathsf{U}_2)\Rightarrow R\leq I(\mathsf{U}_1;\mathsf{U}_2),
\end{align*}
where in the second inequality we have used the assumption that $P_e\to 0$ as $m\to \infty$. This completes the proof.
\begin{Remark}
 Theorems \ref{th:1} and \ref{th:2} provide tight necessary and sufficient conditions except for the case when $R=I(\mathsf{U}_1;\mathsf{U}_2)$. 
\end{Remark}
\section{Special Cases}
\label{Sec:Ex}
In this section, we evaluate the necessary and sufficient conditions for successful matching derived in the previous sections under the I.I.D. and Markovian database models.
\subsection{I.I.D. Database Model}
 A family of pairs of databases generated based on the I.I.D database model is a family of CLDBs  $(\overline{\mathcal{D}}^{(1)}_{m,n_m},\overline{\mathcal{D}}^{(2)}_{m,n_m}), m\in \mathbb{N}$ characterized by  $((n_m)_{m\in \mathbb{N}},(\Theta^{(1)}_m)_{m\in \mathbb{N}}, (\Theta^{(2)}_m)_{m\in \mathbb{N}}, P_{\mathsf{U}^{(1)},\mathsf{U}^{(2)}})$, where $P_{m,\mathsf{U}_i^{(1)}, \mathsf{U}_i^{(2)}}(\prod_{k\in [m]}\mathcal{U}^{1}_k\times \mathcal{U}^{2}_k)=\prod_{k\in [m]} P_{U^{(1)},U^{(2)}}(\mathcal{U}^{1}_k\times \mathcal{U}^{(2)}_k), i\in [n_m], \mathcal{U}^{1}_k\times \mathcal{U}^{(2)}_k\in \mathbb{R}^2$, where $P_{U^{(1)}, U^{(2)}}$ is a probability measure on the two dimensional Euclidean space. Under the I.I.D. model, we have:
\begin{align*}
    I(\mathsf{U}^{(1)};\mathsf{U}^{(2)})&= \lim_{n\to \infty}\frac{1}{n} I(U^{(1),n}; U^{(2),n})
    = \lim_{n\to \infty}\frac{1}{n} \sum_{i\in [n]}I(U^{(1)}_i; U^{(2)}_i)\\
    &= \lim_{n\to \infty}I(U^{(1)}; U^{(2)})
    =I(U^{(1)}; U^{(2)}).
\end{align*}
Consequently, we have the following corollary to Theorem \ref{th:1}.
\begin{Corollary}
For the family of databases generated based on the I.I.D model as described above, a successful matching scheme exists if the following inequality is satisfied:
\begin{align*}
    R< I(U^{(1)}; U^{(2)}),
\end{align*}
where $R$ is the rate of growth of the family of databases. 
\end{Corollary}
\subsection{Markovian Database Model}
A second stochastic database model of interest is the Markovian database model. This model can used for a wide range of databases such as health records and financial transactions where the past elements of an entry are related to the future through the present (and possibly a few of the recent past elements). In the setup described in the previous subsection, assume that the correlated pair of random processes $(\mathsf{U}^{(1)},\mathsf{U}^{(2)})$ are Markov of order $l\in \mathbb{N}$. Then, it is well-known that \begin{align*}
    I(\mathsf{U}^{(1)};\mathsf{U}^{(2)})&=I(U^{(1)}_{l+1}; U^{(2)}_{l+1}|U^{(1),l},U^{(2),l}).
\end{align*}
Consequently, we have the following corollary.
\begin{Corollary}
For the family of databases generated based on the Markovian model as described above, a successful matching scheme exists if the following inequality is satisfied:
\begin{align*}
    R< I(U^{(1)}_{l+1}; U^{(2)}_{l+1}|U^{(1),l},U^{(2),l}),
\end{align*}
where $R$ is the rate of growth of the family of databases. 
\end{Corollary}
\section{Conclusion}
We have investigated the problem of database alignment under a stochastic database model where the correlation among the database entries is governed by arbitrary but known joint distributions. We have used an extension of the Shannon-McMillan-Breiman theorem to propose a database matching scheme. We have leveraged information theoretic tools such as Fano's inequality to provide a converse result which characterizes a set of joint distributions on the database entries for which reliable matching is not possible. We have evaluated the bounds when the database entries are independent and identically distributed and under Markovian database models. 
\label{Sec:Conc}

\bibliographystyle{IEEEtran}
\bibliography{ref}

\end{document}